\documentclass[12pt]{article} 
\usepackage{graphicx}
\usepackage{epsf}
\makeatletter 
\@addtoreset{equation}{section} 
\makeatother 
 
\newcommand{\be}{\begin{equation}}
\newcommand{\ee}{\end{equation}}
\newcommand{\bea}{\begin{eqnarray}}
\newcommand{\eea}{\end{eqnarray}}

\def\dt{\delta}

\def\bb{\bar{\beta}}

\renewcommand{\t}[1]{\tilde{#1}}

\def\Gt{\tilde{G}}
\def\Ph{\varphi}
\def\Pht{\tilde{\varphi}}

\def\Gt{\tilde{G}}
\def\Bt{\tilde{B}}
\def\d{\partial}
\begin{document} 
 
\begin{titlepage}

\begin{center}
\vskip 2.5 cm
{\Large \bf  Non-Abelian T-duality in Pre-Big-Bang Cosmology}
\vskip 1 cm
{\large A. Bossard\footnote{e-mail:\ \tt bossard@celfi.phys.univ-tours.fr}
and N. Mohammedi\footnote{e-mail:\ \tt nouri@celfi.phys.univ-tours.fr}}\\
\vskip 1cm
{\em Laboratoire de Math\'ematiques et Physique Th\'eorique \footnote{CNRS
UPRES-A 6083}\\  Universit\'e Fran\c{c}ois Rabelais \\
Facult\'e des Sciences et Techniques \\
Parc de Grandmont \\
F-37200 Tours - France.}
\end{center}
\vskip 1.5 cm
\noindent

\begin{abstract} 
We study the impact of non-Abelian T-duality transformations on a
string based cosmological model. The implementation of the pre-big-bang
scenario is investigated. We found a region of the dual phase where such a
picture is possible.  
\end{abstract} 
\end{titlepage} 
 
\section{Introduction}
One of the most impressive theoretical constructions of the past century
is undoubtedly the standard cosmological model based on
the general relativistic Friedmann-Lema\^{\i}tre-Robertson-Walker (FLRW)
metric. Not only it does explain the Hubble law,
but it also led Gamov to predict the presence of a
cosmic microwave background (CMB) radiation, detected many years
later. Despite its many successes, such as the explanation of nucleosynthesis, 
this model is not exempt from drawbacks.
Indeed, the standard scenario cannot satisfactorily explain
the observed flatness of space-time nor
the homogeneity of the CMB (the horizon problem)
or the generation of cosmological perturbations
responsible for the big structures of the universe.
In order to remedy to these problems one is forced to assume an era of
formidable expansion of the scale factor of the metric. 
This is the inflation scenario.
However, as argued in \cite{Borde:1994xh}, such an
inflationary phase cannot eliminate the initial big-bang singularity of
standard cosmology.

In order to deal with such a singularity one might
need a quantum description of gravity. Presently, string theory seems
to be the best candidate for describing the gravitational effects of matter
at very small scales. In this theory, fundamental particles
are described by tiny one dimensional strings embedded in external
background fields.
These backgrounds are not chosen arbitrarily.
The internal coherence of string theory imposes constraints
on the background fields, in particular the metric of space-time.
Indeed, one demands that the two dimensional non linear sigma model 
description of the propagation of strings in these backgrounds should be
invariant under Weyl transformations.
This leads to the vanishing of the so called Weyl anomaly
functions (or the generalized beta functions).
This requirement results in a 
generalization of Einstein's equations of general
relativity. Moreover, these Weyl anomaly equations can be viewed as 
equations of motion of an action (the string effective action) containing the
Einstein-Hilbert term.  This action describes the
low energy excitations of string theory (the metric, the dilaton and
the antisymmetric tensor).

Another important property of string theory is a set of transformations known
as T-duality.
These T-dualities relate seemingly two inequivalent sets of backgrounds. 
A string propagating in a space-time endowed with a first background would
present the same dynamics when moving in another space-time endowed with a
dual background. 
The natural area where the effects 
of these dualities might be observed is in cosmology.
In seminal papers \cite{Veneziano:1991ek,pbb1}, Veneziano and Gasperini have
shown that a combination of Abelian T-duality, leading to an inversion of
the scale factor of a flat FLRW metric (scale factor duality), together with a
time reversal operation lead to a phase of inflation.
This inflation is driven by a growing dilaton leading to a strong
gravitational interaction.  
This phase (the dual phase) is valid for negative values
of the time parameter and is called the pre-big-bang phase.
However, the initial big-bang singularity is always present in this theory.
It is sandwitched between two phases: the original flat FLRW solution and its
Abelian dual.

Therefore, a suitable matching of the two phases must be found if one is to
avoid the big-bang singularity. The answer to this problem could be found by
including $\alpha'$ corrections in the string effective action ($\alpha'$ is
the loop parameter of the two dimensional sigma model). The original analyses
of Gasperini and Veneziano were carried out for the lowest order in $\alpha'$.
Higher order contributions in $\alpha'$ might be able to deal with the high
curvature states typical of a big-bang singularity. The problem of
regularization of the big-bang singularity (the graceful exit problem) by
means of higher $\alpha'$ contributions have been theoretically investigated
in  \cite{Gasperini:1996fv,Gasperini:1996fu,Easson:1999xw}
and numerically in \cite{Maharana:1998qc,Chiba:1999jz,Ellis:rw}.
These studies indicate that the pre-big-bang cosmology could provide the
necessary phase of inflation as well as a solution to the pre-big-bang
singularity.  
This theory, however, still has its own drawbacks \cite{Brustein:1994kw,Turner:1997ih}. 

In this paper, we would like to investigate the role of a more complicated
T-duality, namely non-Abelian duality, in cosmology.
Our aim is to figure out if a pre-big-bang picture can be implemented for
a background corresponding to a non flat FLRW space-time.
This space-time presents a natural setting for the application of
non-Abelian T-duality.
We are also motivated by the fact that the string effective action is
invariant under non-Abelian T-duality transformations \cite{Bossard:2000xq}.
We are able to find a restricted region of the dual phase where a 
pre-big-bang approach does indeed apply.
The difficulty in reaching a conclusion regarding the whole dual phase stems
from the absence of any symmetries in the dual backgrounds.
This situation is very typical of non-Abelian T-duality and is due to its
non reversibility \cite{Giveon:1993ai}.

The paper is organized as follows:
In section 2, 
we formulate the general problem and
summarize the main results of non-Abelian T-duality.
In particular, an expression relating the matter content in the original
theory and its dual counterpart is given. This relation is essential for 
the extension of T-duality transformations in the presence of matter fields.
We briefly recall, in section 3, the important features of the
pre-big-bang cosmological model. 
Our example, based on a non flat FLRW metric, is studied in section 4. 
Finally, we present our conclusions and outlook in section 5.

\section{String effective action and T-dualities}
Our starting point is the following model as defined by the action
\bea
S=S_{\rm eff}+S_{\rm mat}\,\,,
\label{5.9}
\eea
where $S_{\rm mat}$ denotes the action for the matter content of some
cosmological model and $S_{\rm eff}$ is the low energy effective action of
string theory given by \bea
S_{\rm eff}=\int d^D X \sqrt{-G} e^{-\Ph}
\Big[R+(\d \Ph)^2-\frac{1}{12}H_{MNP}H^{MNP}+\Lambda \Big]\,\,.
\label{5.10}
\eea
In this expression, $G$ is the determinant of the metric $G_{MN}$ and $R$ is the
associated Ricci scalar.
The field $\Ph$ is the usual dilaton and $H_{MNP}$,
defined by $H_{MNP}=\d_{M} B_{NP}+\d_{N} B_{PM}+\d_{P} B_{MN}$,
is the torsion of the antisymmetric field $B_{MN}$.
Finally, $\Lambda$ stands for a possible cosmological constant.
In the absence of matter fields, the equations
of motion of the action $S$ would reduce to the vanishing of the Weyl anomaly coefficients
$\bar{\beta}^{(\omega)}$ defined by
\bea
\bb^{(G)}_{MN} & \equiv & R_{MN}+\nabla _{M}\partial _{N}\varphi
-\frac{1}{4}H_{MPQ}{H_N}^{PQ} \,\, , \nonumber \\
\bb^{(B)}_{MN} & \equiv & -\frac{1}{2}\nabla ^{P}H_{MNP}
+\frac{1}{2}H_{MNP}\partial ^{P}\varphi \,\, ,\nonumber \\
\bb^{(\Ph)} & \equiv & -\frac{1}{4}\nabla ^{2}\varphi
+\frac{1}{4}\partial _{P}\varphi \partial ^{P}\varphi
-\frac{1}{24}H_{MNP}H^{MNP}
-\frac{\Lambda}{4} \; .
\label{5.11}
\eea
In the presence of matter, however, the equations of motion of the action $S$
can be cast in the following form 
\bea
\bb^{(G)}_{MN} &=& \frac{1}{2}G_{MN}E^{(\Ph)}+e^{\Ph}T^{(G)}_{MN}\,\,,
\nonumber \\
\bb^{(B)}_{MN} &=& e^{\Ph}T^{(B)}_{MN}\,\,,
\nonumber \\
E^{(\Ph)} &=& e^{\Ph}T^{(\Ph)}\,\,,
\label{5.12}
\eea
where 
\bea
E^{(\Ph)} &\equiv& \bb^{(G)}_{PQ} G^{PQ} -4 \bb^{(\Ph)}\,\,,
\nonumber \\
T^{(G)}_{MN} &\equiv& - \frac{1}{\sqrt{-G}} \frac{\dt S_{\rm mat}}{\dt
G^{MN}}\,\,, \nonumber \\
T^{(B)}_{MN} &\equiv&  G_{MP} G_{NQ} \frac{1}{\sqrt{-G}} \frac{\dt
S_{\rm mat}}{\dt B_{PQ}}\,\,, \nonumber \\
T^{(\Ph)} &\equiv& \frac{1}{\sqrt{-G}} \frac{\dt S_{\rm mat}}{\dt \Ph} \; .
\label{5.13}
\eea
We have allowed here the matter to depend on the dilaton as well as the
antisymmetric tensor field.
This is in order to have the most general setting for our scenario. 

Let us now briefly recall the T-duality transformations present in the string
effective action $S_{\rm eff}$.
The origin of these transformations resides in the two dimensional non-linear
sigma model description of string theory \cite{Buscher:sk,Rocek:1991ps,delaOssa:1992vc}.
We will consider here the simplest version of these transformations.
In order to do so, the target space coordinates $X^M$ are split as 
$(x^{\mu},y^i)$ where we assume that an isometry group acts on the
coordinates $y^i$ only.
The dependence on $y^i$ of the metric $G_{MN}$ and the antisymmetric tensor
$B_{MN}$ appears only through the one form $e_i^a dy^i$. 
In other words, our backgrounds are such that 
\bea
G_{\mu \nu} &=&  G_{\mu \nu}(x) \; ,\nonumber\\
B_{\mu \nu} &=&  B_{\mu \nu}(x) \; ,\nonumber\\
G_{\mu i} &=&  Z_{\mu a}(x)e^a_i(y) \; ,\nonumber\\
B_{\mu i} &=&  W_{\mu a}(x)e^a_i(y) \; ,\nonumber\\
G_{ij} &=&  e^a_i(y)S_{ab}(x)e^b_j(y) \; ,\nonumber\\
B_{ij} &=&  e^a_i(y)v_{ab}(x)e^b_j(y) \; ,\nonumber\\
\Ph &=& \Ph(x) \;,
\eea
Here, $e_i^a(y)$ are vielbeins satisfying  
\begin{equation}
\partial _{i}e_{j}^{a}-\partial _{j}e_{i}^{a}=-f_{bc}^{a}e_{i}^{b}e_{j}^{c}
\end{equation} 
and $f^a_{bc}$ are the structure constants of the
non-Abelian isometry group acting on the string backgrounds.

With these assumptions, the string effective action $S_{\rm eff}$ remains
invariant when the original backgrounds
$\omega=(G_{MN},B_{MN},\Ph)$ are replaced by their duals
$\tilde{\omega}=(\Gt_{MN},\Bt_{MN},\Pht)$ \cite{Bossard:2000xq}. The latter
are given by 
\bea
\tilde{G}_{\mu \nu } & = & G_{\mu \nu }
-\bar{S}^{ab}(Z_{\mu a}Z_{\nu b}-W_{\mu a}W_{\nu b})
+\bar{A}^{ab}(Z_{\mu a}W_{\nu b}-W_{\mu a}Z_{\nu b}) \nonumber \\ 
\tilde{B}_{\mu \nu } & = & B_{\mu \nu}
-\bar{A}^{ab}(Z_{\mu a}Z_{\nu b}-W_{\mu a}W_{\nu b})
+\bar{S}^{ab}(Z_{\mu a}W_{\nu b}-W_{\mu a}Z_{\nu b}) \nonumber \\
\tilde{G}_{\mu i} & = & 
-(Z_{\mu a}\bar{A}^{ab}+W_{\mu a}\bar{S}^{ab})\eta _{bi}\nonumber \\ 
\tilde{B}_{\mu i} & = & 
-(Z_{\mu a}\bar{S}^{ab}+W_{\mu a}\bar{A}^{ab})\eta _{bi}\nonumber \\
\tilde{G}_{ij} & = & \eta _{ia}\bar{S}^{ab}\eta _{bj}\nonumber \\ 
\tilde{B}_{ij} & = &\eta _{ia}\bar{A}^{ab}\eta _{bj}\nonumber \\ 
\tilde{\varphi } & = & \varphi -\ln \left(\det M_{ab}\right) \;, 
\eea
where $\eta_{ia}$ is a unit matrix and 
$E_a^i$ are the inverses of the vielbeins $e_a^i$.
Moreover, $\bar{S}^{ab}$ and $\bar{A}^{ab}$ are the symmetric and
antisymmetric parts of the inverse of 
\bea 
M_{ab} \equiv (S_{ab}+v_{ab})  + y^i \eta_{ic}f^c_{ab} \;.
\eea

The non-Abelian T-duality transformations lead also to some
relations between the Weyl anomaly coefficients of the
original backgrounds and the Weyl anomaly coefficients
of the dual backgrounds\footnote{The equations of motion and the Weyl anomaly
functions for the dual theory are obtained from those of the original model
by replacing the untilded quantities by tilded ones.}. 
These are given by
\cite{Bossard:2000xq} \bea \bb^{(\tilde{\omega})}=\sum_{\omega} \frac{\dt
\tilde{\omega}}{\dt \omega}\bb^{(\omega)} \;.
\label{5.14} 
\eea
It is remarkable that the duality transformations leave invariant the quantity 
$E^{(\Ph)}$ \cite{Bossard:2000xq}. Namely, 
\bea
E^{(\Pht)}=E^{(\Ph)}\,\,.
\label{5.15}
\eea
In this paper we will consider exclusively bloc-diagonal
backgrounds ($Z_{\mu a}=W_{\mu a}=0$). The calculation of the partial
derivatives $\dt \tilde{\omega}/\dt \omega$, given at the end of ref.
\cite{Bossard:2000xq}, shows that, in a bloc-diagonal case, the previous
relations (\ref{5.14}) take the form 
\bea
\bar{\beta}^{(\tilde{G})}_{MN} &=& 
\frac{\delta \tilde{G}_{MN}}{\delta G_{PQ}} \bar{\beta}^{(G)}_{PQ} 
+
\frac{\delta \tilde{G}_{MN}}{\delta B_{PQ}} \bar{\beta}^{(B)}_{PQ} 
\;, \nonumber \\
\bar{\beta}^{(\tilde{B})}_{MN} &=& 
\frac{\delta \tilde{B}_{MN}}{\delta G_{PQ}} \bar{\beta}^{(G)}_{PQ} 
+
\frac{\delta \tilde{B}_{MN}}{\delta B_{PQ}} \bar{\beta}^{(B)}_{PQ} 
\;.
\label{5.16}
\eea
Our aim here is to study, for a simple model, non-Abelian duality
transformations in the presence of matter fields.
The demand that these transformations are still a symmetry of the full
action $S$ will require the original matter tensors $T^{(\omega)}$ to
have dual counterparts $T^{(\tilde{\omega})}$.  
Indeed, using the equations of motion of the original and dual models together
with the relations in (\ref{5.15}, \ref{5.16}), one can show 
that 
\bea
e^{\tilde{\varphi}} T^{(\tilde{\varphi})} &=& e^{\varphi} T^{(\varphi)} \;,
\nonumber \\ 
e^{\tilde{\varphi}} T^{(\tilde{G})}_{MN} &=& e^{\varphi} \left[ \frac{1}{2}
\left( \frac{\delta \tilde{G}_{MN}}{\delta G_{PQ}} G_{PQ} -
\tilde{G}_{MN} \right) T^{(\varphi)} 
+ \frac{\delta \tilde{G}_{MN}}{\delta G_{PQ}} T^{(G)}_{PQ}
+ \frac{\delta \tilde{G}_{MN}}{\delta B_{PQ}} T^{(B)}_{PQ} \right]\;,
\nonumber \\
e^{\tilde{\varphi}} T^{(\tilde{B})}_{MN} &=& e^{\varphi} \left[ \frac{1}{2}
\frac{\delta \tilde{B}_{MN}}{\delta G_{PQ}} G_{PQ} T^{(\varphi)}  
+ \frac{\delta \tilde{B}_{MN}}{\delta G_{PQ}} T^{(G)}_{PQ}
+ \frac{\delta \tilde{B}_{MN}}{\delta B_{PQ}} T^{(B)}_{PQ} \right]\;.
\label{5.16.5}
\eea 
The matter sector in the dual theory will be determined
for our cosmological model. Before tackling this issue, we will start by a
short review of pre-big-bang cosmology.

\section{Abelian pre-big-bang cosmology}

The pre-big-bang picture was put forward 
by Veneziano and Gasperini \cite{pbb1,pbb2,Veneziano:1997kx,pbb3}. Its main
ingredient is the use of Abelian T-duality in a string cosmological model.
The space-time is four dimensional with coordinates 
$X^M=(t,y^i)$ where $y^i=(x,y,z)$.
The original backgrounds are chosen to be given by
\bea
G_{MN} &=& {\rm diag}(-1,a^2(t),a^2(t),a^2(t)) \,\,,
\nonumber \\
B_{MN} &=& 0 \,\,,
\nonumber \\
\Ph &=& \Ph(t)\; .
\label{5.17}
\eea
The non-vanishing components of the Weyl anomaly coefficients $\bb^{(\omega)}$
are
\bea
\bb^{(G)}_{00} &=& \ddot{\Ph}-3\dot{H}-3H^2\,\,,
\nonumber \\
\bb^{(G)}_{ij} &=& \dt_{ij} a^2 [\dot{H}+3H^2-H\dot{\Ph}]\,\,,
\label{5.18}
\eea
and
\bea
E^{(\Ph)} &=& 12H^2+6\dot{H}-2\ddot{\Ph}-6H\dot{\Ph}+\dot{\Ph}^2+\Lambda\,\,.
\label{5.19}
\eea
The Hubble function is defined by the usual relation
$H(t) \equiv \frac{d}{dt} \ln a(t)$.

The coupling of the matter in this model is that of a
perfect fluid, as in standard cosmology. 
The matter tensors $T_{MN}^{(\omega)}$ are such that
\bea
T_{MN}^{(G)}&=& {\rm diag}(\rho,pa^2 ,pa^2 ,pa^2 )\,\,,
\nonumber \\
T_{MN}^{(B)}&=&0\,\,,
\nonumber \\
T^{(\Ph)}&=&0\,\,.
\label{5.20}
\eea
The equations of motion  corresponding to the action $S$ are
\bea
\ddot{\Ph}-3\dot{H}-3H^2 &=& e^{\Ph} \rho\,\,,
\nonumber \\
\dot{H}+3H^2-H\dot{\Ph} &=& e^{\Ph}p \,\,,
\nonumber \\
12H^2+6\dot{H}-2\ddot{\Ph}-6H\dot{\Ph}+\dot{\Ph}^2+\Lambda &=& 0\,\,.
\label{5.21}
\eea
The original backgrounds in (\ref{5.17}) possess three translation symmetries
(Abelian isometries). 
The application of Abelian T-duality transformations \cite{Buscher:sk} leads to
the following  dual backgrounds 
\bea
\tilde{G}_{MN} &=&{\rm diag}(-1,a^{-2}(t),a^{-2}(t),a^{-2}(t))\,\,, \nonumber
\\ \tilde{B}_{MN} &=& 0\,\,,
\nonumber \\
\Pht &=& \Ph(t)-6\ln a(t)
\label{5.22}	
\eea
The corresponding non-vanishing dual Weyl anomaly coefficients
$\bar{\beta}^{(\tilde{\omega})}$ are as follows 
\bea
\bb^{(\Gt)}_{00} &=& \ddot{\Ph}-3\dot{H}-3H^2\,\,,
\nonumber \\
\bb^{(\Gt)}_{ij} &=& -\dt_{ij} a^{-2} [\dot{H}+3H^2-H\dot{\Ph}]\,\,,
\label{5.23}
\eea
and
\bea
E^{(\Pht)} &=&12H^2+6\dot{H}-2\ddot{\Ph}-6H\dot{\Ph}+\dot{\Ph}^2+\Lambda\,\,.
\label{5.24}
\eea
If we now suppose, in analogy with the original model, that the matter fields
in the dual model are such that
\bea
T^{(\Pht)}&=&0\,\,,
\nonumber \\
T_{MN}^{(\Bt)}&=&0\,\,,
\nonumber \\
T_{MN}^{(\Gt)}&=& {\rm diag}(\t{\rho},\t{p} a^{-2},\t{p} a^{-2},\t{p} a^{-2})
\;, \label{5.25}
\eea
then the equations of motion of the dual theory are
\bea
\ddot{\Ph}-3\dot{H}-3H^2 &=& e^{\Pht} \t{\rho}\,\,,
\nonumber \\
\dot{H}+3H^2-H\dot{\Ph} &=& -e^{\Pht} \t{p} \,\,,
\nonumber \\
12H^2+6\dot{H}-2\ddot{\Ph}-6H\dot{\Ph}+\dot{\Ph}^2+\Lambda &=& 0\,\,.
\label{5.26}
\eea
These equations are the same as the ones in (\ref{5.21})
provided that the matter fields in the original model and its dual are related
by 
\bea
e^{\Ph} \rho &=& e^{\Pht} \t{\rho} \,\,,
\nonumber \\
e^{\Ph} p &=&  -e^{\Pht} \t{p} \,\,.
\label{5.27}
\eea
Therefore, the duality transformations of $S_{\rm eff}$ can be extended to the
full action $S=S_{\rm eff}+S_{\rm mat}$.
Notice also that if the original matter obeys the equation of state $\rho =
\alpha p$, where $\alpha$ is a constant, then the matter in the dual theory
satisfies the equation of state $\tilde{\rho} = -\alpha \tilde{p}$. We deduce
that if $\alpha$ is positive then the pressure in the dual phase must be
negative.

Let us now examine the cosmological effects of Abelian T-duality. We suppose,
for simplicity, that the dilaton field is constant 
$\varphi=\varphi_0$ and the cosmological constant vanishes. In this case, the
last equation of (\ref{5.21}) leads to $a(t) \sim \sqrt{t}$ while the two
others yield the equation of state $\rho = 3p$. Hence the universe is in a
radiation dominated era. The energy density $\rho$ is related to the scale
factor $a$ by $\rho \sim a^{-4}$.
An important property of this standard
cosmological solution is that of not being defined for all
values of $t$ and has a big-bang singularity at $t=0$.

The idea behind pre-big-bang cosmology is to seek a solution which is defined
for all values of $t$ and avoids the big-bang singularity. This solution is
found through the use of T-duality transformations combined with a time
reversal operation. Indeed, the equations of motion of the original
model (\ref{5.21}) as well as those of the dual theory (\ref{5.26}) are
form invariant under $t \rightarrow -t$.  

The pre-big-bang picture can be summarized by the graph in Figure 1.
\begin{center}
\begin{figure}
\hspace{3cm}
\fbox{\scalebox{0.5}{\epsfbox{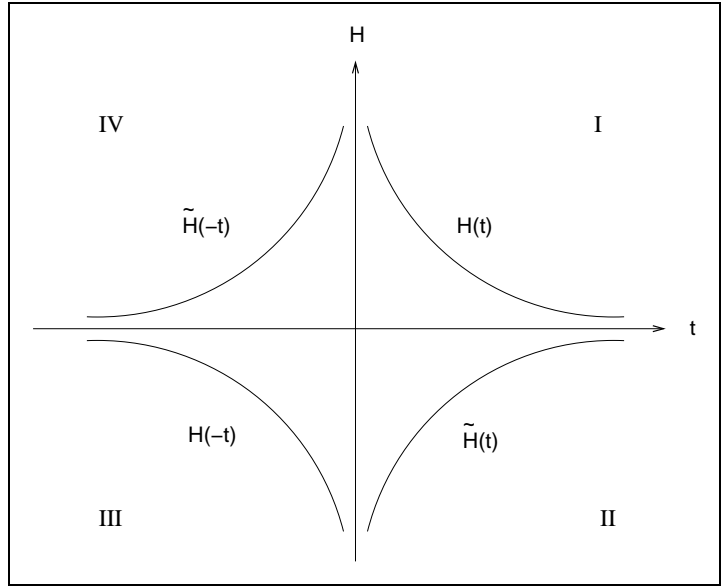}}}
\caption{Abelian T-duality and time reversal}
\end{figure}
\end{center}
One starts from the original solution (region I) as defined by
\bea
a &\sim& \sqrt{t} \; , \nonumber \\
\rho &=& 3p \sim a^2 \; , \nonumber \\
\varphi &=& \varphi_0  
\label{5.?}
\eea
and constructs region II by applying a T-duality transformation.
Region VI is reached by a time reversal operation on the solution
of region II. Therefore, one obtains for region IV the following solution
\bea
\tilde{a} &\sim& \frac{1}{\sqrt{-t}} \; , \nonumber \\
\tilde{\rho} &=& -3\tilde{p} \sim \tilde{a}^{-2} \; , \nonumber \\
\tilde{\varphi} &=& \varphi_0 - 3 \ln (-t) ;.
\label{5.28}
\eea 
This solution is defined for negative times and a full account of its
characteristics can be found in \cite{pbb1,pbb2,pbb3,Veneziano:1997kx}.

Gasperini and Veneziano suggested that by taking region IV when $t$ is
negative and region I when $t$ is positive, one might avoid the big-bang
singularity. However, this requires a non trivial mechanism for matching
these two branches. We refer the reader to
\cite{Gasperini:1996fv,Gasperini:1996fu,Easson:1999xw,Maharana:1998qc,Chiba:1999jz,Ellis:rw}
 for more details on this subject.

\section{Non-Abelian T-duality in pre-big-bang cosmology}

Our goal in this article is to apply non-Abelian T-duality transformations to a
cosmological model with a pre-big-bang scenario in sight.
This is motivated by the fact that most of the interesting cosmological
metrics present, in a natural way, non-Abelian isometries.  
We therefore consider a four dimensional FLRW space-time
described by the coordinates $X^M=(t,y^i)$ with spherical spatial coordinates
$y^i=(R,\Theta,\Phi)$. The string metric is described by   
\bea G_{MN} &=& {\rm diag}
\left(-1,\frac{a^2(t)}{1-kR^2},a^2(t)R^2,a^2(t)R^2\sin^2(\Theta) \right) \; . 
\label{5.29}
\eea
The other string background fields are given by
\bea 
B_{MN} = 0 \; ,\; \;
\Ph = \Ph_0\; .
\label{5.30}
\eea
Notice that the spatially flat case, corresponding to $k=0$, is the one treated
in the previous section. We will assume from now on that $k$ is different from
zero.

In order to find the dual backgrounds, one has to cast the
metric in a form suitable for the application of the formalism of
non-Abelian T-duality. Therefore, one must determine the vielbeins of the
spatial metric.
We distinguish two cases. The first corresponds to $k$ negative.
This case is identified with the isotropic Bianchi V space. The Lie
algebra of the isometry transformations of Bianchi V are known to have
tracefull structure constants $f^a_{ab} \neq 0$. It is precisely for this kind
of Lie algebras that non-Abelian T-duality does not apply.  
This problem with Bianchi V was first revealed in \cite{Gasperini:1993nz}
and further examined in \cite{Tyurin:1994cd,Alvarez:1994zr,Elitzur:1994ri,Bossard:2000xq}.
For this reason, this case will not be considered any further.

The second case deals with $k$ positive.   
This corresponds to the isotropic Bianchi IX space whose isometry Lie algebra
is $SU(2)$ with $f^a_{ab}=0$. Since there is only one
maximally symmetric three dimensional metric with a constant positive
curvature, a change of  coordinates can be found \cite{Weinberg} to bring the
metric in (\ref{5.29}) into the form
\bea ds^2 &=& -dt^2 + h(t) {\rm Tr} \left[ \left(g^{-1} dg\right) \left(g^{-1}
dg\right) \right]\;, \nonumber \\           &=& -dt^2+h(t)\dt_{ab} e^a_i e^b_j
du^i du^j \; , \nonumber \\
&=& -dt^2+h(t)\Big( du^2+dv^2+dw^2+2 \cos(v) du dw \Big)\;. 
\label{5.32}
\eea
In this expression, $g$ is an element of the Lie group $SU(2)$. It is
parametrized by $g=e^{uT_3}e^{vT_1}e^{wT_3}$, where $T_a$ are related to
the usual Pauli matrices $\sigma_a$ by $T_a={\sigma}_a/2i$.
The function $h(t)$ is defined by $h(t)=a^2(t)/4k$ and the
vielbeins are given by 
\bea  
e^a_{\;i}(y)=\left( \begin{array}{ccc}
\sin (v)\sin (w) & \cos (w) & 0\\
\sin (v)\cos (w) & -\sin (w) & 0\\
\cos (v) & 0 & 1
\end{array}\right)\; .
\label{5.31}
\eea
The indices $a$ and $i$ denote, respectively, the lines and rows of $e^a_i$.

To simplify matters, we assume a vanishing cosmological constant ($\Lambda=0$).
These backgrounds leads to 
\bea
E^{(\varphi)} = 6 \left [ \frac{\ddot{a}}{a} + \left (\frac{\dot{a}}{a} \right
)^2 + \frac{k}{a^2} \right] \;.
\label{5.33}
\eea 
The other non-vanishing Weyl anomaly function $\bar{\beta}^{(\omega)}$ 
are
\bea
\bar{\beta}^{(G)}_{00} &=& -3 \frac{\ddot{a}}{a} \;, \nonumber \\
\bar{\beta}^{(G)}_{ij} &=& G_{ij} F(t) \;, \nonumber \\
 F(t)                  &\equiv& 
\ddot{a}/a + 2\left(\dot{a}/a \right )^2 + 2k/a^2 \;.
\label{5.34}
\eea
We also take the matter present in the universe to be 
described by the energy-momentum tensor 
\bea
T^{(G)}_{MN} = 
\left (
\begin{array}{cc}
 \rho & 0 \\
0 & p G_{ij} 
\end{array}
\right ) \;, \; \; T^{(\varphi)} = T^{(B)}_{MN} = 0 \; .
\label{5.35}
\eea 
The equations of motion are then  
\bea
E^{(\varphi)} &=& 0 \;, \nonumber \\
-3 \frac{\ddot{a}}{a} &=&
 e^{\varphi_0} \rho \;,
\nonumber \\
F(t) &=& e^{\varphi_0} p \; , 
\label{5.36}
\eea
Combining these three equations leads to the equation of states
$\rho = 3 p$. Hence, as in the Abelian case, the matter is in a radiation form.
One can
then show that $C=(e^{\varphi_0} \rho a^4 )/3$ is constant in time 
and the general solution for the equations of motion is
given by 
\bea
a(t) = \sqrt{C'}\left[ 1 - \left(1 - \frac{t}{\sqrt{C'}} \right)^2
\right]^{\frac{1}{2}}\;,
\label{5.37}
\eea   
where $C' = C/k^2$.
The graph of $a(t)$ is shown in Figure 2. This solution has a
big-bang singularity at $t=0$ and a big-crunch singularity at $t=2\sqrt{C'}$. 

\begin{center}
\begin{figure}
\hspace{3cm}
\fbox{\scalebox{0.5}{\epsfbox{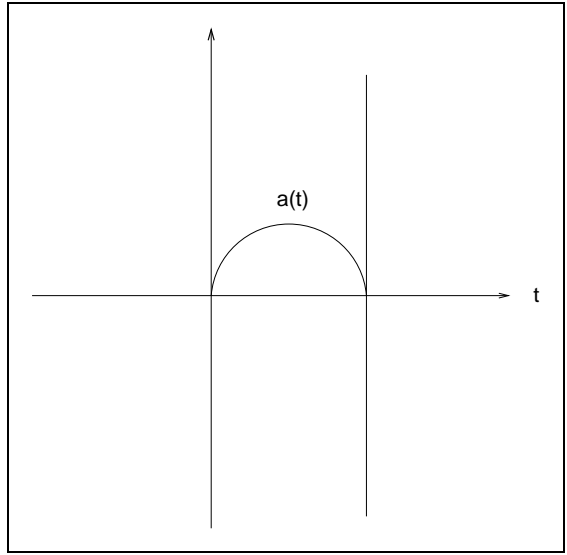}}}
\caption{ Big-bang and big-crunch singularities}
\end{figure}
\end{center}

Now we are in a position to construct the dual backgrounds as explained in 
the second section of this paper.  
One obtains, after a coordinate transformation from
$(t,u,v,w)$ to $(t,r,\theta,\phi)$ \cite{Balog:1998br}, the following dual 
backgrounds\footnote{The change of variables is given by $u=r \sin(\theta)
\cos(\phi),\;v=r\sin(\theta)\sin(\phi),\;w=r\cos(\theta)$.}  

\bea
\tilde{G}_{MN} &=&
{\rm diag}
\left(-1,\frac{1}{h(t)},\frac{h(t)r^2}{\Delta},\frac{h(t)r^2\sin^2(\theta)}{\Delta}\right)
\;, \nonumber \\ 
\tilde{B}_{MN} &=& 
\frac{r^3 \sin(\theta)}{\Delta}\left( \begin{array}{cccc}
0 & 0 & 0 & 0\\
0 & 0 & 0 & 0\\
0 & 0 & 0 & -1\\
0 & 0 & 1 & 0\\
\end{array}\right) \;, \nonumber \\
\tilde{\varphi} &=& \varphi(t) - \ln (h \Delta) \;,
\label{5.38} 
\eea
where $\Delta = h^2(t) +r^2$.
One can see that the metric of this dual backgrounds is still isotropic, due to
the presence of the element $d \theta^2 + \sin^2(\theta) d \phi^2$
in the expression of $d s^2$. But it is not homogeneous anymore.
This kind of metric must then have a {\em center}.

The non-vanishing Weyl anomaly coefficients $\bar{\beta}^{(\tilde{\omega})}$ of
these dual backgrounds are 
\bea
E^{(\tilde{\varphi})} &=& E^{(\varphi)} \; , \nonumber \\
\bar{\beta}^{(\tilde{B})}_{23} &=&
2 r^3 \sin(\theta) \frac{h^2}{\Delta^2}F(t) \; , \nonumber \\
\bar{\beta}^{(\tilde{G})}_{00} &=&
\frac{3C}{(4k)^2}\frac{1}{h^2}\; , \nonumber \\
\bar{\beta}^{(\tilde{G})}_{11} &=&
-\frac{F(t)}{h} \; , \nonumber \\
\bar{\beta}^{(\tilde{G})}_{22} &=&
-\frac{r^2 h (h^2 -r^2)}{\Delta^2}F(t) \; , \nonumber \\
\bar{\beta}^{(\tilde{G})}_{33} &=&
-\frac{r^2 \sin^2(\theta)h (h^2 - r^2)}{\Delta^2}F(t)
\; , 
\label{5.39}
\eea 
By injecting these expressions in the equations of motion of the dual theory,
we are led to the following form for the dual energy-momentum tensor
\bea
T_{MN}^{(\tilde{G})} = 
\left( \begin{array}{cccc}
\tilde{\rho} & 0 & 0 & 0\\
0 & \tilde{p}_1 \tilde{G}_{11} & 0 & 0\\
0 & 0 &  \tilde{p}_2 \tilde{G}_{22} & 0\\
0 & 0 & 0 & \tilde{p}_3 \tilde{G}_{33}\\
\end{array}\right) \;.
\label{5.40}
\eea 
An explicit calculation yields the following expressions for the energy
density and pressures appearing in $T^{(\tilde{G})}_{MN}$
\bea
\tilde{\rho} &=& 3 K \frac{h^2 + r^2}{h} \;, \nonumber \\
\tilde{p}_1 &=& -K \frac{h^2 + r^2}{h} \;, \nonumber \\
\tilde{p}_2 &=& -K \frac{h^2 - r^2}{h} \;, \nonumber \\
\tilde{p}_3 &=&  \tilde{p}_2 \;,
\label{5.41}
\eea
where $K$ is defined by $K= e^{-\varphi_0}C/(4k)^2$.
Furthermore, the equations of motion of the dual theory are fulfilled only if 
\bea
T^{(\tilde{\varphi})} &=& 0 \;, \nonumber \\
T_{23}^{(\tilde{B})} &=& -2K \tilde{B}_{23} h \;.
\label{5.42}
\eea
This means that in the dual phase, the matter depends on the dual
antisymmetric tensor $\tilde{B}_{MN}$. This is to be contrasted with the
original theory where we have set $B_{MN}=0$ (implying $T^{(B)}_{MN}=0$).
The non-vanishing of $\tilde{B}_{MN}$ and $T^{(\tilde{B})}_{MN}$
breaks the isotropy observed in the dual metric $\tilde{G}_{MN}$.
We should mention that the expressions of $T^{(\tilde{G})}_{MN}$,
$T^{(\tilde{B})}_{MN}$ and $T^{(\tilde{\varphi})}$ are compatible with the
formal relations in (\ref{5.16.5}).

\section{Discussion}

Let us now investigate whether one can set a pre-big-bang scenario based on
the original backgrounds and their duals. As before, the equations of motion
are still form invariant under time reversal $t \rightarrow -t$.
Our original solution has two singularities and the pre-big-bang picture,
based on non-Abelian duality and time reversal transformations, 
would be expected to avoid these singularities.
This task, however, is difficult to achieve due to the fact that the dual
backgrounds do not carry the symmetries of their original counterparts.
Nonetheless, some conclusions can be reached in some limiting procedure. Two
cases, depending on the relative values of $r^2$ and $h^2$, emerge
naturally in the study of the dual backgrounds.

The first of these two cases corresponds to the situation when $r^2<<h^2$.
The dual backgrounds tend then to the following expressions
\bea
\tilde{G}_{MN} & \rightarrow & {\rm
diag}(-1,\frac{1}{h},\frac{r^2}{h},\frac{r^2 \sin^2(\theta)}{h}) \;, \nonumber
\\ \tilde{B}_{MN} & \rightarrow & 0 \;,
\nonumber \\
\tilde{\varphi} & \rightarrow & \varphi_0 - 3 \ln h \;.
\label{5.42.5}
\eea
Similarly, the dual matter fields are described by 
\bea
\tilde{\rho} & \rightarrow & 3Kh \;, \nonumber \\
\tilde{p}_1 & \rightarrow & \tilde{p} = -Kh \;, \nonumber \\
\tilde{p}_2 & \rightarrow & \tilde{p} = -Kh \;, \nonumber \\
\tilde{p}_3 & \rightarrow & \tilde{p} = -Kh \;, \nonumber \\
T^{(\tilde{B})}_{23} & \rightarrow & 0 \;.
\label{5.43}
\eea
In this limiting zone, the universe is described by a flat FLRW metric with
a scale factor given by $1/\sqrt{h(t)}$. Furthermore, the only matter present
is that of a perfect fluid obeying the equation of state
$\tilde{\rho}=-3\tilde{p}$. One can, therefore,
define a Hubble function $\tilde{H}= - \frac{d}{dt} \ln \sqrt{h}$.

In analogy with the Abelian pre-big-bang scenario, we construct in this limit a
similar setting as represented in Figure 3.
The Hubble function of region I is that of the original theory, while region
IV corresponds to the metric in (\ref{5.42.5}) but for negative $t$.
 \begin{center}
\begin{figure}
\hspace{3cm}
\fbox{\scalebox{0.5}{\epsfbox{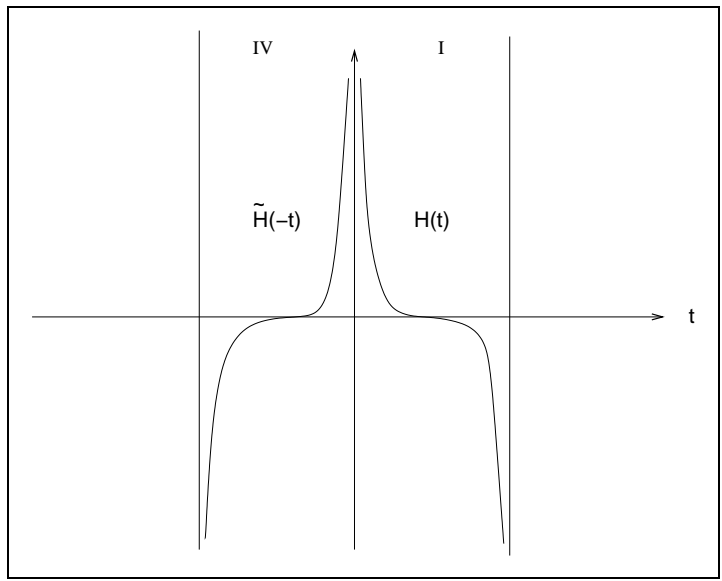}}}
\caption{Non-Abelian T-duality and time reversal} 
\end{figure}
\end{center}
As one can see, the behaviors of $H(t)$ and $\tilde{H}(-t)$ are similar to
what happened in the Abelian case (see Fig.1) near the big-bang singularity.
Here again, one must solve the graceful exit problem (the matching of the two
branches). There are, however, two other singularities (big-crunch) when    
$t \rightarrow \pm 2\sqrt{C'}$ (see Fig. 2). We speculate that higher order
corrections in $\alpha'$ might cure the model of all these singularities.

The other limiting case corresponds to $r^2>>h^2$.
Although, we are not able to construct a pre-big-bang cosmology for this
limit, its study reveals some interesting features. 
The dual backgrounds in this zone tend to
\bea
\tilde{G}_{MN} & \rightarrow & {\rm diag}(-1,\frac{1}{h},h,h
\sin^2(\theta)) \;, \nonumber \\
\tilde{B}_{MN} & \rightarrow & r \sin(\theta)\;,
\nonumber \\
\tilde{\varphi} & \rightarrow & \varphi_0 - \ln (h r^2) \;.
\eea
On the other hand, the dual matter is represented by the quantities
\bea
\tilde{\rho} & \rightarrow & 3K\frac{r^2}{h} \;, \nonumber \\
\tilde{p}_1 & \rightarrow &  -K \frac{r^2}{h}\;, \nonumber \\
\tilde{p}_2 & \rightarrow &  K \frac{r^2}{h}\;, \nonumber \\
\tilde{p}_3 & \rightarrow &  K \frac{r^2}{h}\;, \nonumber \\
T^{(\tilde{B})}_{23} & \rightarrow & -2Khr \sin(\theta) \; .
\eea
We notice that the $r$ coordinate gets a special status in these backgrounds.
The dual metric contains a sphere of radius $\sqrt{h}$ and whose surface is
described by the coordinates $\theta$ and $\phi$. When $h$ is big the $r$
direction shrinks while the sphere expands. The opposite happens for small
$h$. This situation is similar to the procedure of dynamical compactification
encountered in Kaluza-Klein cosmologies \cite{Chodos:vk}.
Furthermore, the pressure along the $r$ direction is negative.

In conclusion, we have managed to put forward a pre-big-bang
cosmological scenario in a very particular region ($r^2<<h^2$) in the dual
phase. However, a global study of the full dual phase seems at the moment out
of reach. The lack of symmetries in the dual backgrounds (absence of isotropy
or homogeneity) does not allow a construction of a Hubble function. Therefore,
a comparison with the original FLRW backgrounds is not, in general, possible.

Finally, we should mention that we have analysed another
cosmological model based on Poisson-Lie T-duality. Unfortunately, the dual
backgrounds are even more involved and do not reveal any known features.
This shows all the difficulties whenever one deviates from the flat case of
Abelian T-duality (scale factor duality). As already remarked in
\cite{Feinstein:1997js,Feinstein:2001ta}, this is very puzzling from a string
point of view: a fundamental string does not see any differences between
two backgrounds related by any T-duality transformations. Therefore, the role
of isotropy and homogeneity in string cosmology is certainly worth
investigating.

\end{document}